\documentclass{aastex62}

\received{January 1, 2018}
\revised{\today}
\accepted{April 22, 2018}
\submitjournal{ApJ}

\shorttitle{Electron-Positron pairs creation close to a black hole horizon.}
\shortauthors{Laurent \& Titarchuk}

\begin{document}

\title{\bf Electron-Positron pairs creation close to a black hole
horizon. Red-shifted  annihilation line in the emergent X-ray spectra of a black hole, part I }

\correspondingauthor{L. Titarchuk}
\email{titarchuk@fe.infn.it}

\author{Philippe  Laurent}
\affiliation{CEA/DRF/IRFU/DAp, CEA Saclay, 91191 Gif sur Yvette, France; philippe.laurent@cea.fr}
\affiliation{Laboratoire APC, 10, rue Alice Domon et L\'eonie Duquet, 75205 Paris Cedex 13, France}

\author{ Lev Titarchuk}

\affiliation{Dipartimento di Fisica, Universit\`a di Ferrara, Via Saragat 1, I-44100 Ferrara, Italy;  titarchuk@fe.infn.it}

\begin{abstract}
We consider a Compton cloud (CC) surrounding a black hole (BH)  in an accreting black hole system, where electrons propagate with thermal and bulk velocities.
In that cloud, soft (disk)  photons may be upscattered off  these energetic electrons and attain several MeV energies. They could then create pairs due to these photon-photon interactions. 
In this paper, we  study  the formation of  the 511 keV annihilation line due to this photon-photon interaction, which results in the creation of electron-positron pairs, followed by the annihilation of the created positrons with the CC electrons.  The appropriate  conditions of  annihilation  line  generation take place very close to  a BH horizon  within $(10^3-10^4)m$ cm  from it,  where $m$ is the BH hole mass in solar units. As a result,   the created annihilation line should be seen by the Earth observer  as a blackbody bump, or so called  {reflection}  bump at  energies around $(511/20) (20/z)$ keV where $z\sim 20$ is a typical gravitational red-shift experienced by the created annihilation line photons when they emerge. This transient feature should occur in any accreting black hole systems, galactic or extragalactic. Observational evidences for this feature in several  galactic black hole systems is detailed in an accompanying paper (II). An extended hard tail of the spectrum up to 1 MeV may be also formed due to X-ray photon upscattering off  created pairs. 
\end{abstract}


\section {Introduction}

Accreting stellar-mass black holes (BHs) in Galactic binaries exhibit high-soft and low-hard spectral states and transitions between them (the intermediate state; see details in \cite{st09}, hereafter ST09).  Usually an increase in the soft blackbody (BB) luminosity component leads to the appearance of an extended power law. However, the extension of the power law is not a monotonic function of the power law index. Using data of the {\it Rossi} X-ray Time Explorer ({\it RXTE})
for black hole candidate (BHC) XTE  J1550-564  \cite{ts10} demonstrate that efold energy $E_{fold}$ of the spectra decreases when the photon index $\Gamma$  increases starting from $\Gamma \sim 1.4$   and then, at some value of $\Gamma\sim 2.2$, it starts increasing towards energy of 200 keV.    An important observational fact is that this effect is seen as a persistent phenomenon only in BH candidates, and thus it is apparently a { unique} BH signature. In neutron star (NS) sources, $E_{fold}-$energy gradually decreases towards the soft state [see e.g. \cite{seit11} and \cite{tsf13}, hereafter TSF13]. Although in NSs, similar power law components are detected in the hard and intermediate states (hereafter the LHS and the IS, respectively)  their extension becomes smaller (or  $E_{fold}$)  with increasing luminosity  [see \cite{DiSalvo00} and \cite{ft11}, hereafter FT11].

Moreover, ST09 find that photon index $\Gamma$ saturates with the dimensionless mass accretion rate $\dot m=\dot M/ \dot M_{crit}$ where $\dot M_{crit}=L_{\rm Edd}/c^2$, with $L_{\rm Edd}=4\pi GMm_p c/\sigma_{\rm T}$, is the Eddington luminosity, $m_p$ the proton mass, $\sigma_{\rm T}$ the Thomson cross-section and $c$ the speed of light. Naturally, the photon index $\Gamma$   saturates  with a quasi-periodic oscillation frequency  (QPO) (and the mass accretion rate) in BH sources while FT11 and TSF13 demonstrate that $\Gamma$ stays almost the same around 2 in NSs. Thus,  it seems a reasonable assumption that  we deal here with  a unique spectral signature  of BH binaries which is directly tied to the black hole event horizon.

In \cite{lt99} (hereafter LT99), we suggested that the BH X-ray spectrum in the high soft state (HSS) is formed in a relatively
cold accretion flow with a relativistic bulk velocity $v_{bulk} \sim c$ and
with  the plasma  (electron) temperature of a few keV or less,  for which the thermal electron velocity $v_{th}\ll c$. It is worth  pointing out that  in such a flow
the effect of the bulk Comptonization is much stronger  than the effect of the thermal Comptonization which
is a second order with respect to $v_{th}/c$ (LT99).  Very close to the horizon,  X-ray photons can be up-scattered by bulk electrons to very high energies, of order  1 MeV and higher but the observer on Earth will see these up-scattered photons at lower energies, of the order of 300 -- 400 keV, because of gravitational redshift.  

Using Monte Carlo simulations, LT99 compute    X-ray spectra of such BH sources. They take into account up-scattering of the soft (disk) photons illuminating the Compton cloud (CC) and they implement  the full relativistic treatment to reproduce these spectra.
The resulting  spectra obtained using this treatment can be described as a sum of a thermal (disk) component and the convolution of some fraction of this component with the CC upscattering spread (Green's) function. The upscattered part of the spectrum is seen   as an extended power law over energies much higher than the characteristic energy of the soft photons. 
LT99 show that the photon index $\Gamma$  increases with an increase of the mass accretion rate  
and then it stabilizes at $\Gamma=2.8\pm0.1$, similarly to what was observed by ST09. This index  stability  occurs  
over a wide range of the plasma temperature, 0 -- 10 keV, and dimensionless mass accretion rates   $\dot m$ (higher than 2 in Eddington units).  

LT99 also demonstrate that the sharp high-energy cutoff occurs at energies of 200-400 keV, which are related to the average energy of electrons, $m_ec^2$ impinging on the event horizon.  Although, the spectrum is practically identical to the standard thermal Comptonization spectrum [see e.g. \cite{ht95}] when the CC plasma temperature is getting of order of 50 keV (the typical ones for the hard state of BH). In this case, one can only see the effect of the bulk motion  at high energies, where there is an excess in the spectrum with respect to the pure thermal one. Furthermore, LT99 demonstrate that the change of spectral shape from the soft X-ray state to the hard X-ray state is clearly to be related to the Compton cloud temperature.  Indeed, the effect of the bulk Comptonization compared with the thermal one is getting stronger when the plasma temperature drops below 10 keV. 

Furthermore, \cite{lt01}, hereafter LT01,  show that the high-energy photon production (source function) is distributed with the characteristic maximum at about the photon bending radius $1.5r_{\rm S}$, where $r_{\rm S}=2GM/c^2$ is the Schwarzschild radius and $M$ is a BH mass, independently of the seed (soft) photon distribution. 
Most of these photons fall down then into the black hole, but
some of them anyway have time to interact with another X-ray photon
by photon-photon process to make an
electron-positron pairs \citep{sven82}.  In this paper, we explore in details the
consequences of this pair creation process which occurs very close  to  a BH horizon,   $(10^3-10^4)m$ cm  from it ($m$ is the BH hole mass in solar units)
 and we elaborate the observational consequences  of this effect.

In the next section, we proceed with the details of the Monte Carlo simulations of  the upscattering  of the soft (disk) photons in the converging flow, and of the pair creation due to the interaction of these up-scattered photons with unscattered X-ray photons in the close vicinity of a BH horizon. Then, we explore in details the observational consequences of this effect. In the Appendix, we give an analytical derivation of some key results obtained through the simulations.

\section{The Monte-Carlo simulations}
The geometry used in these simulations is similar to the one used in
LT99, consisting of
a thin disk  with an inner radius of 3$r_{\rm S}$,  
merged with  a spherical Compton Cloud (CC) harboring a BH in
its center.  The CC outer radius is $r_{out}$. The disk is
assumed always  to be optically thick.
In addition to the free-fall into the central BH,   with a bulk
velocity of the infalling plasma given by $v(r)=c(r_{\rm S}/r)^{1/2}$, where $r_{\rm S}$ is  the Schwarzschild radius
(\cite{tmk97}), we have also taken into account the
thermal motion of the CC electrons, simulated in most of the results
shown here at an electron temperature of 5 keV, 
a typical temperature of the CC in the high-soft state of
galactic BHs (see Fig. \ref{geometry}). The CC  temperature is  less than  the temperature of the sub-Keplerian inflow (presumably located above
the disk) but it is higher than the temperature of the disk flow
which is  of order of a few keV or less (e.g. \cite{ss73}).

The seed X-ray photons were generated uniformly and isotropically
 at the surface of the border of the accretion disk,
 from $r_{d,in}= {3} r_{\rm S}$ to $r_{d, out} = 10r_{\rm S}$.
 These photons were generated according to a thermal spectrum with a
single temperature of 0.9 keV, similar to the   ones measured in
black hole binary systems (see for example \cite{BOR99}). To reach a satisfying statistic level for our computations, we made simulations using 10$^9$ seed (disk) photons.

 In Figure \ref{geometry} we present a geometry of X-ray spectral formation in BH source. Soft X-ray photons coming from the disk illuminate 
the Compton cloud  and its innermost part, where the bulk velocity is dominating in the converging flow (CF) region). These photons upscatter off electrons in these configurations and some part of these photons come to the observer,  seen as a specific Comptonization spectrum.

In our simulations [see also the kinetic formalism in  \cite{tz98}, hereafter TZ98] we use
 the number density  $n_{ff}$ of the flow measured in
the laboratory  frame of the flow  
$n_{ff}=\dot m(r_{\rm S}/r)^{1/2}/(2R\sigma_T)$. Here $\dot m=\dot M/\dot M_{Edd}$,
$\dot M$ is  the mass accretion rate, $\sigma_T $ is the Thomson cross section,
and  $\dot M_E \equiv L_{\rm Edd}/c^2=4\pi GMm_p/ \sigma_Tc~$ is
the Eddington accretion rate.
 The parameters of
our simulations are thus, the BH mass $m$ in  solar units, the CC electron
temperature, $T_e$, the mass accretion rate,  $\dot m$, 
and the CC outer radius, $r_{out}=10 r_{\rm S}$.
The  cloud Thomson optical thickness 
is expressed  through  the mass accretion rate $\dot m$ according to the
following formulae (see e.g. TZ98) :

\begin{equation}
\tau_{ff} = \frac{\dot m}{2}\int_{1}^{x_{out}}\frac{dx}{x^{3/2}\sqrt{1-x^{-1}}}
=\dot m \left( \frac{\pi}{2} - \arcsin{\sqrt{r_s \over r_{out}}} \right )
\label{tau_lab}
\end{equation}
\noindent
where $x_{out}=r_{out}/r_{\rm S}$.

In Figure \ref{ener_shift}, we show the average photon energy of the  upscattered photons (black histogram) and blue-shifted energy of injected soft photon (blue line) as a function of radius. As one can see,  the injected (disk) photons illuminating  in the CF innermost region are strongly blue-shifted; at 30 meters  from a BH horizon (for a 10 solar masses BH) their energy is of order of few MeV.
One should keep in mind however that only a few of these photons can directly reach a BH horizon when the accretion rate is high (in the HSS) as the CF optical depth $\tau_{ff}\gg 1$ for $\dot m\gg1$ (see Eq. \ref{tau_lab}).
In addition to the Compton effect, we compute the probability of making a pair, using the pair creation cross section given by \cite{sven82}. The kinetic properties of the pair are computed in the local rest frame using standard formulae, derived from the energy and momenta conservation, and the results stored in a file for a subsequent treatment. The created pair properties are described below.
Afterward, using another Monte-Carlo simulation, we propagate the pairs in the Schwarzschild metric, slowing them down by Coulomb scattering, and checking if the positrons annihilate or not. The resulting annihilation photons were also tracked by a third Monte-Carlo in the CC, to see if they escape or are scattered off due to the CC optical depth.

\section{Created pairs properties}

As one can  see from Fig. \ref{DensityProfile}, the pairs are
produced in the Monte-Carlo simulations very close to  a BH horizon. For a ten solar masses black hole, this creation occurs even at a few hundred meters away from the horizon.
This result  is confirmed by  computations given in Appendix.

The pair creation process depends on the properties of the
background X-ray photons, and, in particular, of the source X-ray
luminosity. Then  we have  simulated the pairs creation process for
different value of the source X-ray  luminosity and computed for 
each of them  the number of positrons created close to the black hole. These
numbers are presented  in Figs. $\ref{DensityLumlh}$-$\ref{DensityLumhs}$ for the  
low/hard  and the high/soft states, respectively.  We choose  these two BH spectral
states in order   
to demonstrate a trend in the creation  of positron number as a function of the luminosity for
two cases of the model parameters, the plasma temperature, $kT_e$ and $\dot m$, even if these states does not span on their own  the whole luminosity range.  We should  point out   that although  the two curves are  similar each other,
the low-hard state (LHS) plot is situated a factor five below the 
high-soft state (HSS) one. This is mainly due to the difference in the  mass accretion rate $\dot m$ between these  two cases. This linear dependence of the number of the created pairs  on $\dot m$   occurs when the luminosity reaches $10^{36}$ erg/s, when the number of created positrons is equal to
around $10^{5} L_X$ and  $5\times 10^{5} L_X$  for the LH and
HS states, respectively (see Figures \ref{DensityLumlh}$-$\ref{DensityLumhs}).
Simulations also  show  that, independently
of the source X-ray luminosity, the positron kinetic energy
spectrum remains nearly the same.

\section{The emergent photon spectrum of an accreting black hole}
\subsection{A redshifted 511 keV annihilation line feature}

As we have already discussed, the pair creation  occurs very close
to the horizon, of the order of   100 meters  from the the horizon for a 10 solar mass BH.
In Figure \ref{trajectoire}, we show the
trajectories of the created positrons in the Schwartzchild
background. It can be seen that only a few of them, among the 5x10$^5$ positrons simulated to produce this Figure, can reach $3~r_{\rm S}$. This number is even
lower if we take into account Coulomb losses and  the pair annihilation process. 

Having the results of our first Monte-Carlo, we have built another Monte-Carlo software with which we simulate the propagation of the created 
positrons in the cloud, taking into account energy
losses due to Coulomb scattering with protons. This process is the
dominant cooling one, considering  the physical condition of the
cloud (see discussion in \cite{MJ97}). We compute
also the annihilation probability for positrons, and check for each
positron if it annihilates or not. If the annihilation occurs then  we store the kinetic
energy and location of this positron at the moment of the
annihilation. If not, these positrons can either fall into a BH, Compton scatter off background photons, 
 escape from the system or annihilate further away, outside the CC, depending on the density of the
surrounding material at  several Schwartzchild radii. If we suppose
that all positrons escaping from the CC annihilate afterward, the
maximum expected 511 keV photon flux  is around $10^{36}$ photon/s (i.e. $10^{30}$ erg/s) for a
luminosity of $10^{38}$ erg/s. 
In a further step, we derive, using a third Monte-Carlo code and the saved annihilated positrons locations, the emergent annihilation radiation. 
As the 511 keV annihilation emission is produced at different radius from
the BH, the line endures different redshifts, depending on the location of the annihilation
and  thus,  this annihilation process shows up to the external observer as a specific continuum. 
In Figure \ref{anni5}, we show the resulting spectra for
a CC electron  temperature of 5 keV. In this set of simulations,  $\dot m$ 
was set equal to 1 (blue line), 4 (red line), and 10 (green line).
The spectrum become harder as $\dot m$ increases. Also, due to the concurring effects of the pair creation efficiency, which increases with $\dot m$, and CC opacity to the 511 keV radiation, decreasing with $\dot m$, we could see in Figure \ref{anni5} that the spectrum is maximum for intermediate values of $\dot m=4$.

Then we have  explored what happens if we keep constant  the mass accretion rate  ($\dot m = 4$), 
while varying the CC electron temperature. The result  can
be seen  in  Figure \ref{anni50}, where we present the emergent
annihilation spectrum with a CC electron  temperature of 5 keV (blue line) and
50 keV (red line). 
One can notice that the overall spectrum shape does not change
much except that the spectrum shift toward higher energies when we
increase the cloud temperature, $kT_e$.

In Figure \ref{anni5_total}, we show also the emergent Comptonized spectrum of the CC, with an electron temperature of $kT_e=5$ keV, including the gravitationally redshifted  annihilated line.  On the upper encapsulated right-hand panel, we include the redshifted  annihilation line as a ratio of the resulting spectral values  to the Comptonized continuum.  These simulations show that the emergent annihilation spectrum is quasi-universal whatever the CC physical
conditions are.
In Figures \ref{anni5}-\ref{anni50},   we  demonstrate also that these redshifted  annihilation lines  can be represented by a blackbody-like shape, as shown by the dash lines which gives the blackbody best fit for the $\dot m = 1$ and $\dot m = 4$ case respectively.

However, observationnally, this feature should be transient and cannot be observed in all black hole spectral states. Indeed, if the resulting  X-ray luminosities $L_{\gamma} \gg  10^{37}(M_{bh}/10 M_{\odot})$ erg/s, the dimensionless mass accretion rate (see \cite{tz98}) is much greater than 1, and the produced pairs and  annihilation photons generated near a BH horizon  cannot escape, as they are effectively scattered off converging electrons.  In the LHS we cannot see  the HBB either, and consequently  the redshifted annihilation line, because  $\tau_{\gamma - \gamma} \ll 1 $ and pairs are not effectively generated. Moreover, in LHS, the converging flow is surrounded by a hot, relatively thick  Compton Cloud (CC) with plasma temperature  of order of 50-80 keV. Thus, the escaping  HBB photons are scattered off hot electrons of the CC and the bump is smeared out. So, we expect this feature to be seen only during the IS when the mass accretion rate is sufficient to trigger the effect but no strong enough to hide it afterward.

\subsection{Emergent spectrum taking into account the pair creation}
In the previous section, we demonstrated that the pairs are created very close to horizon  due to photon-photon interactions. In  fact, these photons reach the necessary conditions for the pair creation when the product of their  energies $E_1E_2>(m_ec^2)^2$. In Figure \ref{addspec}, we show 
the redshifted annihilation line and the classical bulk motion Comptonizarion  spectrum,  plus the result of the photon upscattering off these energetic pairs which leads to an extension of the emergent spectrum up to 10 MeV. In that case, the CC temperature is 5 keV ($\dot m = 4$). 
In Figures \ref{lena}$-$\ref{EnergySpectrum} we also demonstrate the general relativity (GR)  geometry,   photon and pair distributions (spectra)   for diifferent radii within  the converging flow, respectively.

It is worth noting that similar spectra in the  HSS  in quite a few BH binaries  were observed by  the OSSE and COMPTEL instruments [see  \cite{grove98} and \cite{Mc94}, respectively]. In particular, seven BH transient sources, GRO J0422+32, GX 339-4, GRS 1716-249, GRS 1009-45, 4U 1543-47, GRO J1655-40, and GRS 1915+105, observed 
by OSSE demonstrated two Gamma spectral states (HSS and LHS) and the transition between these two states.  \cite{grove98} emphasize that, in the LHS, the emergent spectra are characterized by hard spectra with photon indices $\Gamma<2$ and exponential cut off  of energies $\sim 100$ keV. 
This form of the spectra is consistent with the thermal Comptonization case  [see \cite{st80}]. These spectra were observed by \cite{grove98} for  GRO J0422+32, GX 339-4, GRS 1716-249, GRS 1009-45.  While  GRS 1009-45, 4U 1543-47, GRO J1655-40, and GRS 1915+105 show the high soft state (HSS)  with relatively soft photon index, $\Gamma$  in the range from 2.5 to 3. Furthermore, \cite{grove98} have found that the HSS emergent spectrum in   GRO J1655-40 is extended to 690 keV without any sign of the rollover at low energies. One can compare this Grove's spectrum with  our Monte Carlo simulated  HSS spectrum   demonstrated in Figure \ref{addspec}.  

\cite{Mc94} provided details of the Cyg X-1 observations in the 0.75-30 MeV energy range using the COMPTEL instrument in CGRO. They  found  the Cyg X-1 spectrum in the HSS to be extended  at least up to 1 MeV without any break. No physical explanation  was suggested to explain this observational result besides a suggestion of incorporating  a spectral component which represents the reflection of the hard X-rays from an optically thick accretion disk (see \cite{hm93}).   However, this reflection scenario is inconsistent with the soft state emergent Cyg X-1 spectrum because spectra with photon index higher than 2 do not demonstrate the reflection effect. Indeed, the spectrum is too steep and there is no enough photons at high energies to transfer them to lower energies to make a reflection (or downscattering) bump [see \cite{lt07}].   However, we argue using our simulations that the combined OSSE-COMPTEL spectrum for Cyg X-1 in the HSS demonstrated in \cite{Mc94} can result from the Comptonization of the soft (disk) photons by pairs in the converging flow, as shown in Fig. \ref{addspec}.      

\section{A redshifted annihilation radiation. A high temperature BB  (HBB) Component}

\cite{TSC18}, hereafter TSC18 (paper II) found observational evidences of the high temperature BB (HBB)  bump around 20 keV, which could be fitted by a $\sim$ 4.5 keV blackbody  profile, in black hole accreting systems.  
Evolutions of spectral characteristics of GRS 1915+105, SS~433 and V4641 Sgr  are very similar and 
all of them  reveal the  HBB feature which is  centered around   20 
keV. Using {\it RXTE} and  INTEGRAL data, TSC18  demonstrated that the HBB component is needed in the IS broad-band spectra  of SS 433.
On the other hand the comparative analysis of the {\it Beppo}SAX and {\it RXTE} data
shows that flaring pattern for GRS~1915+105 and V4641~Sgr,  respectively,
demonstrates a transition from  the IS to the LHS.
It is interesting   that the HBB feature is usually observed in the same 
X-ray luminosity range ($5\times 10^{36}$ -- $5\times 10^{37}$ erg/s) for different sources. 
Thus, one  can raise a natural question on the origin of this HBB.
 
 TSC18 estimated the optical depth for photon-photon interaction  very close to a BH horizon. In order to do this,  they  calculated the photon density $N_{\gamma}$ near a black hole horizon, assuming that most photons have energy greater than $m_ec^2=511$ keV:  
\begin{equation}
N_{\gamma}=\frac{L_{\gamma}}{4 \pi r^2 c m_e c^2},
\label{phot_dens}
\end{equation}   
where $L_{\gamma}\simeq 10^{37}(m/10)$ erg/s, $ r_{\rm S} = 2GM/c^2$ is
the Schwarzschild radius,  (or $r_{\rm S} = 3 \times 10^6(m/10)$ cm), 
$c$ is the speed of light ($3\times 10^{10}$ cm/s) and 
the electron rest energy, $m_e c^2$ is about $5\times 10^{-7}$ erg. As a result, they obtain that $N_{\gamma}=0.6\times 10^{19}/(M_{bh}/10 M_{\odot})$ cm$^{-3}$. 
Then the optical depth for photon-photon interactions can be estimated as 
\begin{equation}
\tau_{\gamma - \gamma} \sim \sigma_{\gamma - \gamma} N_{\gamma} r_{\rm S}.
\label{tau_phot_phot}
\end{equation}  
The cross section for  $\gamma-\gamma$ interaction,  $\sigma_{\gamma - \gamma}\sim 0.2\sigma_T$, where $\sigma_T=6\times 10^{-25}$ cm$^2$ is 
the Thomson  cross-section. 
Thus, they find  that $\tau_{\gamma - \gamma} \sim (1 - 2)$ using 
Eqs. (\ref{phot_dens}--\ref{tau_phot_phot}).  It is important to emphasize  that $\tau_{\gamma - \gamma}$ is independent of the BH mass.
We demonstrate in our Monte Carlo  simulations (see also TSC18), that the IS luminosity of order $10^{37}
$ erg s$^{-1}$  is sufficient to get  a high enough optical depth $\tau_{\gamma - \gamma}$ to trigger photon-photon interaction very close to a BH horizon.

So, pairs (electrons and positrons) are effectively  generated as a result of $\gamma - \gamma$ interactions
near a BH horizon, within a shell of order of 100 $(M_{bh}/10 M_{\odot})$ m. The generated positrons propagate and may interact with accreting electrons leading to the formation of the annihilation line at 511 keV. A significant fraction of these line photons can directly escape to the Earth observer if the Klein-Nishina optical depth, $\tau_{\rm KN}$ at 511 keV is of order of one. In the way out these 511 keV line photons undergo gravititational redshift with $z$ around 20 forming the HBB with color temperature around 20 keV.  Observationally this HBB feature has an equivalent width in the interval  from 400 to 700 eV (see TSC18, Fig. 9).

As noticed in \S 4, in the HSS, the dimensionless mass accretion rate is much greater than 1, and the produced pairs and  annihilation photons generated near a BH horizon  cannot escape, as they are effectively scattered off converging electrons. In the other hand, in the LHS, the converging flow is surrounded by a hot, relatively thick  Compton Cloud (CC) with plasma temperature  of order of 50-80 keV, and the escaping  HBB photons are scattered off hot electrons of the CC and the bump is smeared out. So, it is not  by chance that  we cannot observe the HBB bump in the HSS when $\dot m$ is high  (see TSC18, Fig. 9 there) neither in the LHS,  but only in the transient IS, where $\dot m$ is around 1 -- 2.  In Fig. \ref{EWmdot},  we plot the simulated bump equivalent width evolution with $\dot m$.

\section{Conclusions}
We made extensive Monte Carlo simulations of  X-ray spectral formation in the cloud surrounding a black hole. We found the emergent spectrum extends to relatively high energies up to 400 keV due to up upscattering of the soft (disk) photons illuminating the CC region. Moreover, one can observe a redshifted annihilation line formed near a black hole horizon when the  dimensionless mass accretion rate $\dot m$ is in the range of 1 -- 2 and the plasma temperature $kT_e$ is of order or lower than 5 keV.  These conditions are met when a  BH is observed in the intermediate state. 
The annihilation line is formed due the pair production when the upscattered photon energies exceed 511 keV threshold.  The generated positrons in their way out  annihilate with incoming electrons produce the 511 keV line. This line being produced very close to a BH horizon, the photons undergoes a significant redshift of order of 20 while going out to the Earth observer and the line is observed as an high energy bump around 20 keV.

This annihilation line cannot be observed in the  LHS characterized by a low mass accretion rate,  less than one,  and a hot Compton cloud with plasma temperature about 50 keV surrounding the converging flow site (see   Figs. \ref{geometry} and \ref{EWmdot}).        
Moreover, in the HSS, the annihilation line is effectively generated very close to a horizon but the line photons cannot escape as they are scattered in the optically thick converging flow (see  Fig. \ref{EWmdot}). This line is then observed in the IS, as it demonstrated in the accompanying paper (TSC18).

When we take into account the non-linear effect of the positron-electron (pair) generation very close to a BH horizon we find that the  surrounding photons  upscattered off generated pairs form an additional hard tail extended up to a few MeV (see Fig. \ref{addspec}).   It is worth noting that the observed spectra  of a few BHs  demonstrated this high energy extension [see \cite{Mc94}, \citep{grove98}]  but up to now there were not any  reasonable explanation of this phenomenon in the literature.   

\begin{acknowledgements}
 We should   acknowledge the referee’s efforts on the clear presentation of our paper.
\end{acknowledgements}


\appendix

\section{Analytical derivation of some key results of the simulations}
\subsection{Elements of General Relativity: Photon trajectories in curved space. Lengths and optical paths}
In the flat space, trajectories of unscattered photons is a straight line for which equation can be written in terms of the sinus theorem:
\begin{equation}
r(1-\mu^2)^{1/2}={\hat p}
\label{str_line}
\end{equation}
where $r$ is the length of a radius vector $\bf r$ at a given point of the line, $\mu=\cos{\beta}$ is the cosine of the zenith  angle $\beta$ between the radius vector ${\bf r}$ and the straight line
 and $\hat p$ is the impact parameter of this line.
Unscattered photon trajectories in the Schwarzschild background is just a generalization of this sinus theorem (TZ98)
\begin{equation}
\frac{x(1-\mu^2)^{1/2}}{(1-1/x)^{1/2}}=p
\label{sch_line}
\end{equation}
where $x=R/R_{\rm S}$ and $R_{\rm S}=2GM/c^2$ is the Schwarzschild radius, $p={\hat p}/R_{\rm S}$ and  $M$ is a BH mass.

Using equation (\ref{sch_line}) we can explain   all properties of the photon trajectory  in the Schwarzschild background. In other words:
\begin{itemize}
\item i. The photon can escape from the black hole horizon to infinity or vice versa for  if $p <p_0=(6.75)^{1/2}$, (see TZ98). 
\item
ii. The photon, for which $p =p_0$,  undergoes circular rotation at  $x=1.5$ (at $3  GM/c^2$). 
\item
iii. All photons for which $p >p_0$ and $x<1.5$   are gravitationally attracted by the black hole. In this case all photon trajectories are finite.
\item
iv. All photons which starts at $x>1.5$ and have $p>p_0$ escape to infinity if they are not scattered off electrons in the way out.
\end{itemize}

The photon trajectory can be presented in polar coordinates $r$ and $\varphi$.  
Using the metric of  the Schwarzschild background we can write (see Fig. \ref{lena} ) that 
\begin{equation}
rd\varphi=\frac{\tan\beta dr}{(1-1/x)^{1/2}}.
\label{dr_dphi}
\end{equation}
Then it follows from Eqs.(\ref{sch_line}-\ref{dr_dphi}) that 
\begin{equation}
\varphi=\int_{1}^x\frac{d\eta}{\eta^2\sqrt{1/p^2-(1-1/\eta)/\eta^2}}.
\label{phi_r}
\end{equation} 
This formula is identical to that derived in  \cite{LL71} using the Hamilton-Jacobi (formalism) equation.
In fact, the equation of the photon trajectory in polar coordinates readily follows from the sinus theorem (Eq. \ref{sch_line}) and the Schwarzschild metric.
Equation (\ref{phi_r}) is valid for any nonzero $p-$photon trajectory except a circular one at $x_0=3/2$ for which $p_0=(6.75)^{1/2}$.  
Formula (\ref{phi_r}) should be numerically calculated. 
From formula (\ref{sch_line}) it is evident that $\mu_0=1$ at $x_0=1$, i.e. the photon always enters to a BH horizon along radial direction.

The length of the radial trajectory($p=0$)   is an integral   
\begin{equation}  
l=r_{\rm S}\int_{1}^x\frac{dx}{(1-1/x)^{1/2}}.
\label{rad_length}
\end{equation}  

For $\alpha =(x-1)\ll 1$  the length  from $r_{\rm S}$ to $r$ is 
\begin{equation}  
l\approx 2r_{\rm S} \alpha^{1/2}.
\label{rad_dif_length_small}
\end{equation} 

For $x\gg1$  we have 
\begin{equation}  
l\approx r_{\rm S}x 
\label{rad_length_large}
\end{equation}

The Thomson  optical path on radial trajectory from $r_{\rm S}$ to $r$ (or from $x=1$ to 
$x=r/r_{\rm S}$) is  
\begin{equation}  
T_{\rm T}(r, r_{\rm S})=r_{\rm S}\int_{1}^x\sigma_{\rm T}n_e(\eta)d\eta,
\label{op_path}
\end{equation}
where $\sigma_{\rm T}$ is the Thomson cross-section and
\begin{equation}
n_e(r)=\dot m(r_{\rm S}/r)^{1/2}/(2r\sigma_{T})
\label{density}
\end{equation}
is the electron density 
[see \cite{tzt02}
for the derivation of $n_e$].
The optical path  can be found analytically as 
\begin{equation}  
T_{\rm T}(r, r_{\rm S})=\tau_{\rm T}(x,1)=\frac{\dot m}{2}\int_{1}^x\frac{dx}{x^{3/2}\sqrt{1-1/x}}=
\dot m [\pi/2-\arcsin(1/x)].
\label{op_path_an_0}
\end{equation}
 
Hence,  the total radial Thomson optical depth of the converging flow is
\begin{equation}  
\tau_0=\tau_{\rm T}(\infty,1)=(\pi/2)\dot m.
\label{tot_op_depth_0}
\end{equation} 
For the Thomson optical paths of the finite trajectories entering to BH horizon for which 
$\alpha=(x-1)\ll 1$ we have 
\begin{equation}
\tau_{\rm T}(r, r_{\rm S}) = \sigma_{T}n(r_{\rm S})l \approx \dot m \alpha^{1/2}. 
\label{tau_small}
\end{equation}

\subsection{Pair creation effect near BH horizon. The photon shell}
If the energetic photons have a high number density $n_{ph}\gg 1/(\sigma_{\rm T}l)$, where $l$ is the charactestic scale of the region within which
the Comptonization (in our case the  Bulk motion Comptonization) takes place, then the process of electron-pair production by two photons becomes important.
Indeed, when a Compton upscattered photon
of energy $E_1$ interacts with another one of energy $E_2$, it may
produce an electron-positron pair provided their energies satisfy the
following relationship:
\begin{equation}
E_1 E_2 >  (m_e c^2 ) ^2. 
\label{pair_con}
\end{equation}
\noindent
The related cross-section for pair production $\gamma_1+\gamma_2\to e^++e^-$ by two photons $\sigma_{\gamma\gamma}$  \citep{ab65}
depends on the product
$y=[E_1E_2/2(m_ec^2)^2](1-{\bf\Omega_1\Omega_2})$, where $(E/c){\bf \Omega}$ is the momentum for a given photon of energy $E$.
The cross-section is only nonzero when $y>1$.  The maximum of  $\sigma_{\gamma\gamma}(y_{max})\approx0.26\sigma_{\rm T}$ takes place at $y^2\approx2$.    

The mean energy of the photons upscattered in the converging flow $<E>$   \citep{tmk97}

\begin{equation}
<E>=E_0[1+4/\dot m]^{N_{sc}}
\label{mean_en}
\end{equation}
This formula is obtained without taking into account the gravitational redshift.
It presumably works in the part  of the converging flow where $(1-1/x)^{1/2}$ is of order of unity (namely 
where $x \gtrsim 1.2$ ). The mean number of scattering $N_{sc}\sim \dot m$. 
Then $<E>/E_0\sim16$ and $<E>=24$ keV  for $\dot m=4$ and $E_0=1.5$ keV.
These photons of energy $<E>$ when they propagate towards the BH horizon are blueshifted.
Their energy increases by a factor 
\begin{equation}
<E>_{bs}/<E>=1/(1-1/x)^{1/2}.
\label{blue_en}
\end{equation}
Photon  energy of order $m_ec^2=511$ keV and higher are achieved in the narrow shell around a BH horizon 
\begin{equation}
0<\alpha=x-1<(<E>/m_ec^2)^2.
\label{width_shell}
\end{equation}
For a given $<E>=24$ keV we obtain that $\alpha<2.5\times10^{-3}$. For 10 solar masses black hole 
the thickness of the shell is about 70 meters where the pairs can be created due to photon-photon 
interactions.
In Figure \ref{ener_shift} we show the average photon energy of Monte Carlo simulated spectrum as a function of radius
in converging flow. We asssume in this Monte Carlo  simulations that the dimensionless accretion rate of the converging flow  
$\dot m=4$ and the electron temperature $kT_e=5$ keV. Energy of the injected soft photons is 1.5 keV.
As seen from  Figure \ref{ener_shift}, the average photon energy is about 25 keV at radius 
$R=1.2 R_{\rm S}$, which is very close to our estimate presented above (see Eq. 
\ref{mean_en}).  Also, we can see from this  Figure  that the blue-shift and Comptonization are equally important below $R=1.02 R_{\rm S}$. 

The photon density in the converging inflow $n_{ph}$ near the  BH horizon 
can be estimated using the photon flux injected in the flow
$F_{inj}=L_{inj}/E_0$.
Thus 
\begin{equation}
n_{ph}\sim \frac{F_{inj}}{4\pi r_{\rm S}^2c}=3\times10^{21}(L_{inj}/10^{37}~\rm erg~s^{-1})(m/10)^{-2}
(E_0/1.5~{\rm keV})^{-1} ~~~~{\rm cm^{-3}}
\label{ph_dens}
\end{equation}      
and the free path for the pair creation is
\begin{equation}
l_{\gamma\gamma}\sim (n_{ph}0.25\sigma_{\rm T})^{-1}= 2\times 10^3 (L_{inj}/10^{37}~{\rm erg~s}^{-1})^{-1}
(m/10)^2(E_0/1.5~{\rm keV})~~~{\rm cm}.
\label{fr_gam}
\end{equation}

The typical length of the photon trajectories $l_{cross}$ in the shell of width $\alpha$ 
can be estimated using formula 
(\ref{rad_dif_length_small}) 
\begin{equation}
l_{cross}=2\alpha^{1/2} R_{\rm S}>3\times10^5(m/10)~~~{\rm cm}
\label{l_cross}
\end{equation}
for $\alpha=2.5\times10^{-3}$.
So the shell photons  have enough time to create the pairs because 
\begin{equation}
t_{cross}/t_{\gamma\gamma}=l_{cross}/l_{\gamma\gamma}\gg1.
\label{time_c_gam}
\end{equation}
Optical depth for the pair creation is
\begin{equation}
\tau_{\gamma\gamma}=l_{cross}/l_{\gamma\gamma}>150(L_{inj}/10^{37}~{\rm erg~s}^{-1})
(m/10)^{-1}(E_0/1.5~{\rm keV})^{-1}.
\label{tau_gam}
\end{equation}

\subsection{The  pair  spectrum}
The spectrum of the pairs created in the photon shell  can be calculated using the shell photon spectrum.
In Figure \ref{spec_rad} we show the Monte-Carlo  photon spectra computed
for three different radius ranges in the flow. 
The black  histogram is the emergent  spectrum seen by observers on Earth.
The green  histogram is the spectrum seen by observers staying at radial distance  $R=(1-1.1)R_{\rm S}$ from the black hole. 
The blue  histogram is the spectrum for observers staying at radial distance  $R=(1-1.01)R_{\rm S}$.
 
As we argue above the pairs have to be created in the shell very close to the horizon. As seen from the shell 
photon spectrum presented in Figure \ref{spec_rad} (blue  histogram there) 
there are a plenty of photons of energy much higher than  511 keV that along with the photons of energies between 10-100 keV  (see a bump in the blue  histogram) can create a noticeable amount of pairs.
The photon spectrum is quite flat, the  index of the high energy power-law part being about 1.2 ($1.2\pm 0.1$). 
It is worth pointing out that the shape of the spectrum is very close to a broken power-law.

Because of the gravitational 
blueshift and upscattering the peak of spectrum is located at energy  about 25 keV that is 17 times more the  energy of the injected
soft photons (1.5 keV).  The photon index of 1.2 is a typical index for the saturation Comptonization.
In fact, the photon index $\Gamma$ can be calculated (see \cite{TL95}, hereafter TL95; \cite{ETC96}) 
\begin{equation}
\Gamma=1+\frac{\ln (1/P)}{\ln(1+\eta)}, 
\label{index_sat}
\end{equation}
where $P$ is the probability of photon scattering in a cloud (shell) and $1+\eta=<E^{\prime}>/E$ is a mean efficiency of 
the photon energy change at any scattering of photon off electrons.
According to  TL95 (in their notation the scattering probability is $\lambda$), 
\begin{equation}
1/P=\exp(\beta)=[\tau\ln(1.53/\tau)]^{-1}.
\label{prob}
\end{equation}
This formula is derived for the plane geometry and $\tau\ll1$.

In the photon horizon shell the bulk velocity of the flow $v$ is very close to the speed of light $v=c(r/r_{\rm S})^{1/2}$.  
For $R=1.01R_{\rm S}$, the Lorentz factor $\gamma=[1-(v/c)^2]^{1/2}=10$.
The mean efficiency (see e.g. \cite{POZ83})
\begin{equation}
<E^{\prime}>/E=1+\eta=[1+\frac{4}{3}(\gamma^2-1)].
\label{effic}
\end{equation}
For $\dot m=4$ the shell optical depth $\tau\approx0.4$ (see Eq. \ref{tau_small}). We assume for this estimate
that $\alpha=(R-R_{\rm S})/R_{\rm S}=0.01$.  We obtain the photon index $\Gamma\sim 1.13$ using formulae (\ref{index_sat}-\ref{effic}) where we put
$\gamma=10$ and $\tau=0.4$.  This value of $\Gamma=1.13$ is very close to that, $\Gamma=1.2\pm 0.1$,  obtained in our MC simulations.

The spectrum of the electrons (positrons), so called the differential yield of the pairs produced by the annihilation of
two photons is proportional to the rate of a number of collisions in a given volume (shell). 
In general terms this rate of number of collisions $\dot N_{col}$ 
is calculated according to \cite{LL71} as follows
\begin{equation}
\dot N_{col}=\dot n_{-}(\gamma_{-}) =C_{f}c\int_0^{\infty}n_{ph}(\epsilon_1)d\epsilon_1\int_{0}^{\infty}n_{ph}(\epsilon_2)\
\sigma(\gamma_{-}, \epsilon_1, \epsilon_2)d\epsilon_2,
\label{yield}
\end{equation}
where $\epsilon=E/m_ec^2$ is the dimensionless photon energy, $\gamma_{-}$, $\gamma_{+}$ are the Lorentz factor of the
electron/positron, $\sigma(\gamma_{-}, \epsilon_1, \epsilon_2)$ is the differential cross-section of the pair production
cross-section and a factor $C_f$ is less than one.

Formula (\ref{yield}) is valid for an isotropic radiation field where 
$\sigma(\gamma_{-}, \epsilon_1, \epsilon_2)$ is the differential cross-section of the pair production
cross-section averaged over a solid angle and $C_f=1/4$, (see \cite{bs97}).  In our MC simulation we implement the exact form of the cross-section where 
the angular dependence has been included. However, we want to demonstrate  
that formula (\ref{yield}) (formally derived for isotropic radiation) can still be applicable for the realistic situation 
modeled in our simulations.

 The exact but quite complicated expression  for  this formula was derived by \cite{bs97}.  
Now, let us consider the simple approximations of formula ({\ref{yield}) and compare them with that obtained in the simulations.

The first approximation of formula (\ref{yield}) is related to a $\delta-$function approximation  of the cross-section 
suggested by \cite{ZL85}, hereafter ZL85:
\begin{equation}
\sigma(\gamma_{-}, \epsilon_1, \epsilon_2)\approx \frac{1}{3}\sigma_{\rm T}\epsilon_2\delta(\frac{\epsilon_1}{2}-\gamma_{-})
\delta(\frac{2}{\epsilon_1}-\epsilon_2).
\label{delta_appr}
\end{equation}
To justify this cross-section approximation ZL85 argue that the photons with $\epsilon_1\gg 1$ will produce $e^{+}e^{-}-$ pairs mostly with $\gamma_{-}\approx \gamma_{+}\approx \epsilon/2$
(\cite{br71}) while colliding with photons of energies of $\epsilon_2\approx2/\epsilon_1$ (\cite{H74}).

Integration in formula (\ref{yield}) using  equation (\ref{delta_appr}) leads to 
\begin{equation}
\dot n_{-}(\gamma_{-}) \propto c\sigma_{\rm T}\frac{1}{\gamma_{-}}n_{ph}(1/\gamma_{-})n_{ph}(2\gamma_{-}).
\label{yield_approx1}
\end{equation}  
$\Gamma_{pair}$ should be around of 2.5 
because the photon index of $n_{ph}(1/\gamma_-)$ is about 0.5  for  $1/\gamma_-\ll 1$  and that of   $n_{ph}(2\gamma_-)$ is around 1(see Fig. \ref{spec_rad}).
In Figure \ref{EnergySpectrum} we present the simulation results for pair energy distribution.
As one can see the simulated  pair spectrum is well approximated by a power law of index 2.5. 

   \begin{figure}
   \centering
\includegraphics[width=14cm, height=14cm]{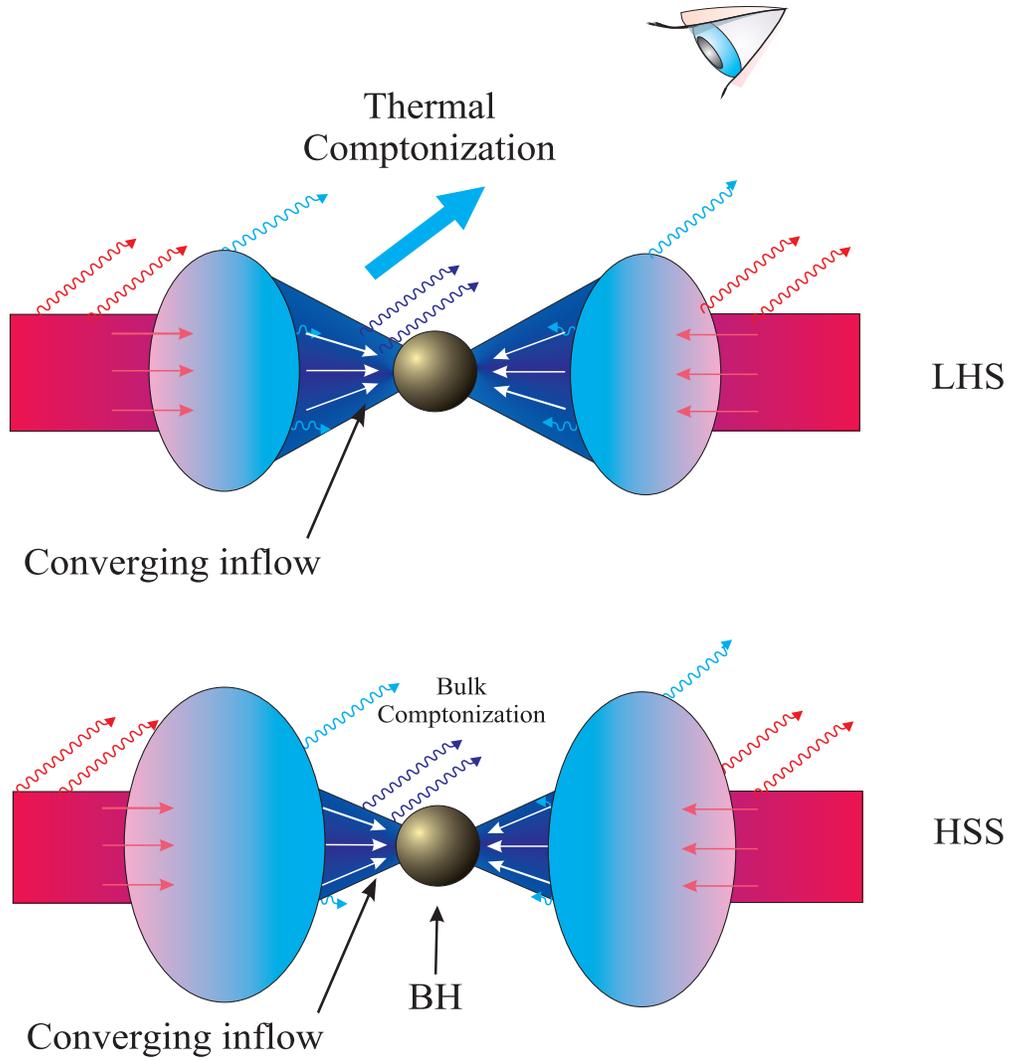}
      \caption{Geometry of a black hole binary. We present two corresponding panels for the low/hard state (LHS) and  for the high/soft state (HSS). 
       }
         \label{geometry}
   \end{figure}

%
   \begin{figure}
   \centering
   \includegraphics[width=10cm, height=10cm]{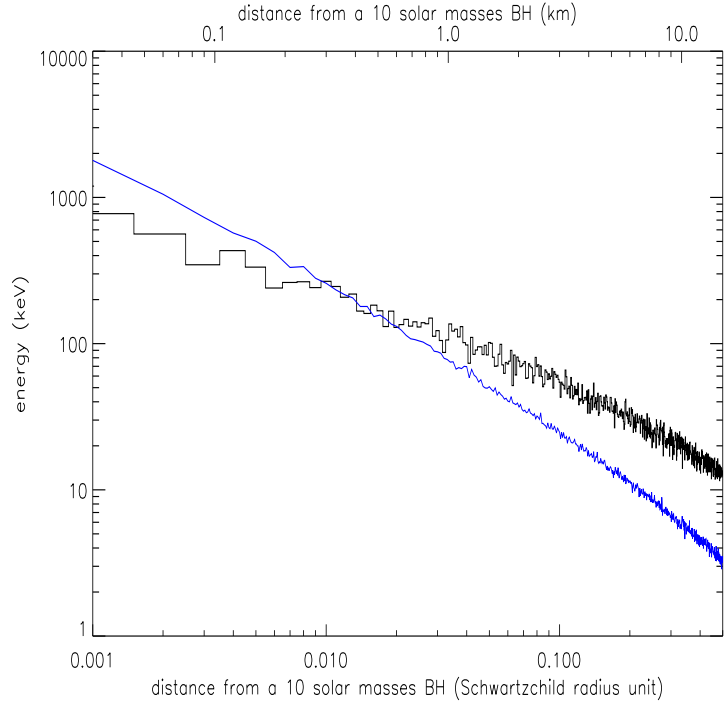}
      \caption{Average photon energy of the upscattered photons (black histogram) and blue-shifted energy of injected soft photon (blue line) as a function of radius.
       }
         \label{ener_shift}
   \end{figure}
%
%
   \begin{figure}
   \centering
   \includegraphics[width=10cm, height=10cm]{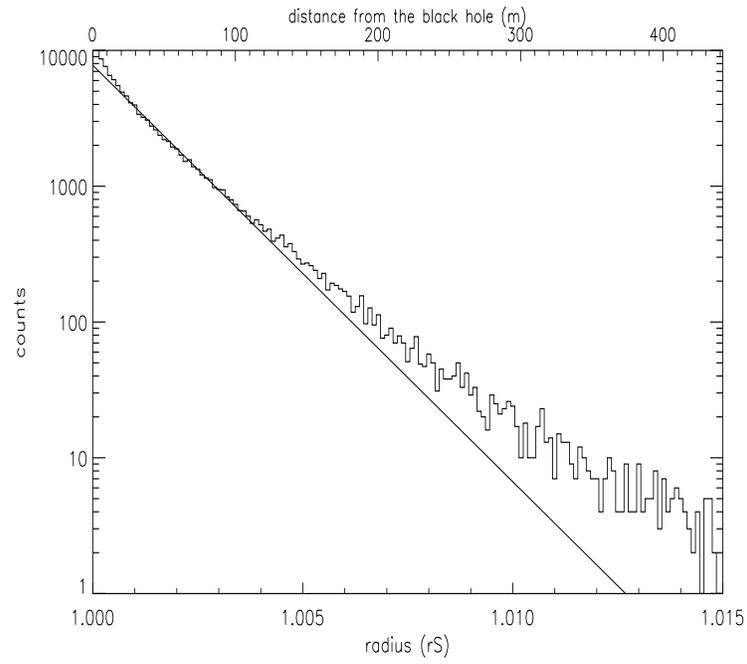}
      \caption{Number density profile of the created pairs as a function of  radius.
       }
         \label{DensityProfile}
   \end{figure}
%
%
%
%
   \begin{figure}
   \centering
   \includegraphics[width=10cm, height=10cm]{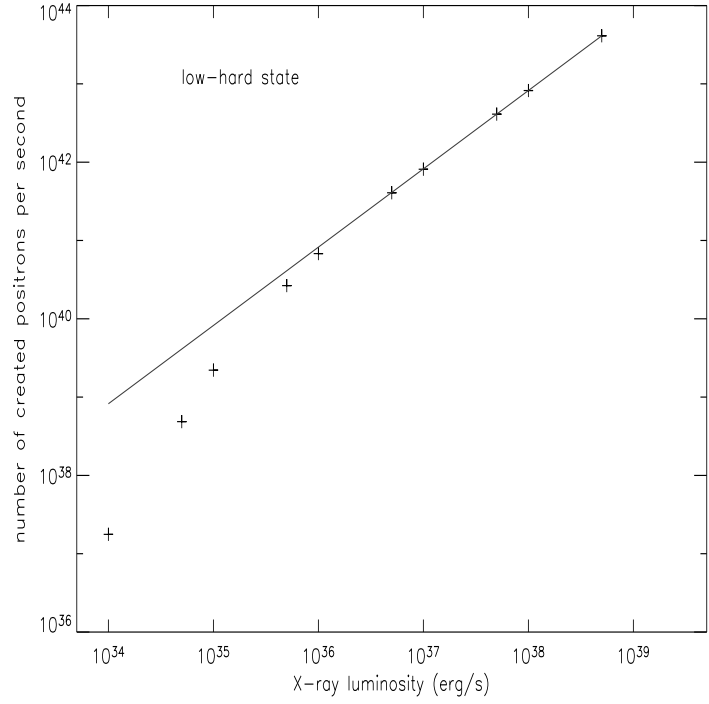}
      \caption{Number of created positrons for a given X-ray luminosity of the source 
      in the low-hard state ($kT_e = 50$ keV, $\dot m = 2$).}
    \label{DensityLumlh}
   \end{figure}
%
   \begin{figure}
   \centering
   \includegraphics[width=10cm, height=10cm]{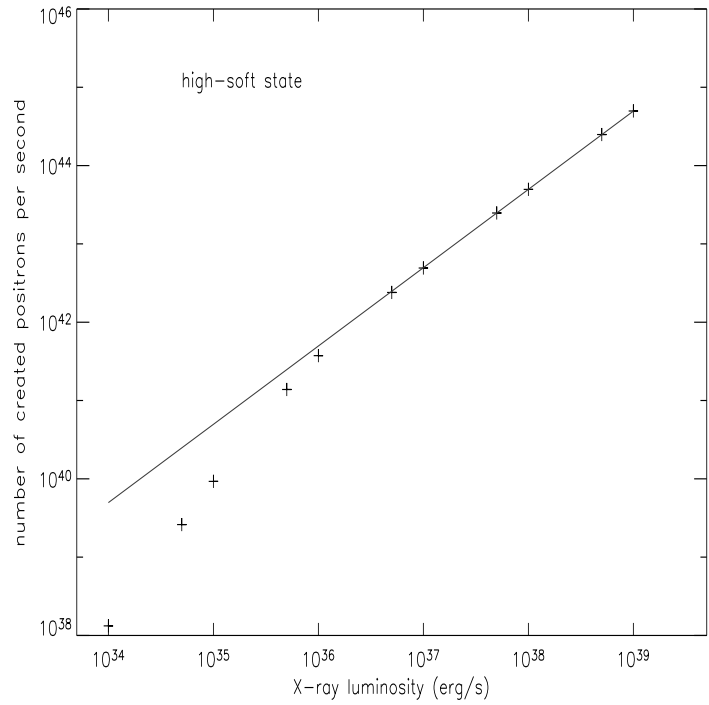}
      \caption{Number of created positrons for a given X-ray luminosity of the source 
      in the high-soft state ($kT_e = 5$ keV, $\dot m = 4$).}
         \label{DensityLumhs}
   \end{figure}

   \begin{figure}
   \centering
   \includegraphics[width=15cm, height=12cm]{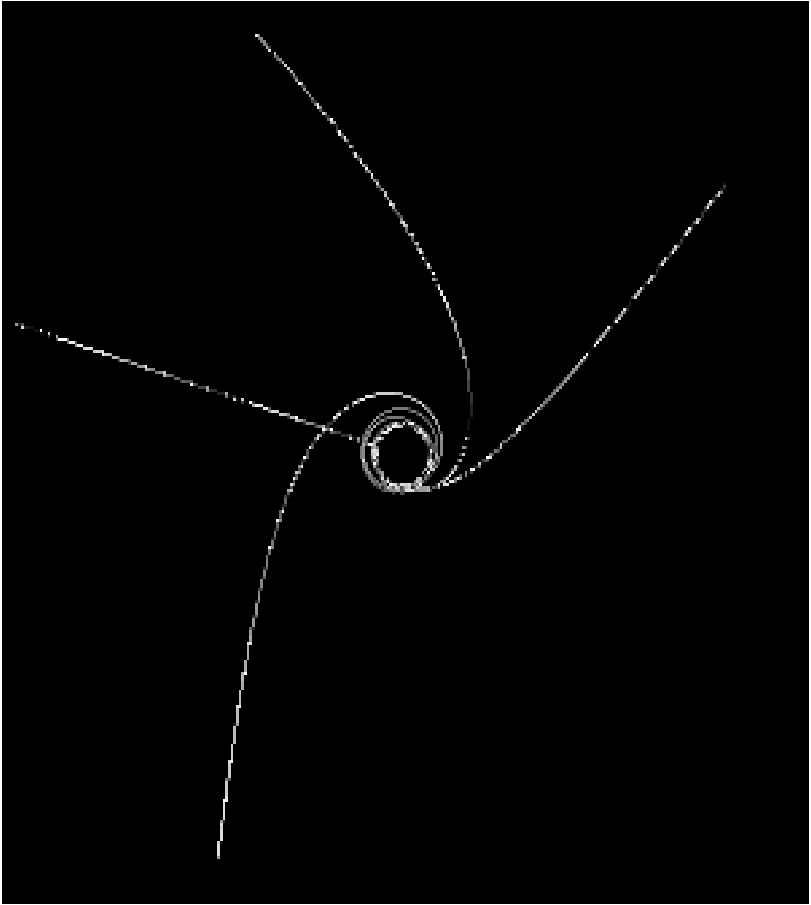}
      \caption{Trajectories of the created positrons in the Schwartzchild background.
      Only a few positrons among the 5x10$^5$ simulated here can reach $3 r_{\rm S}$. This number is even lower if we take into account
      Coulomb losses and annihilation at the positrons (see text)}
         \label{trajectoire}
   \end{figure}
%
   \begin{figure}
   \centering
   \includegraphics[width=15cm, height=12cm]{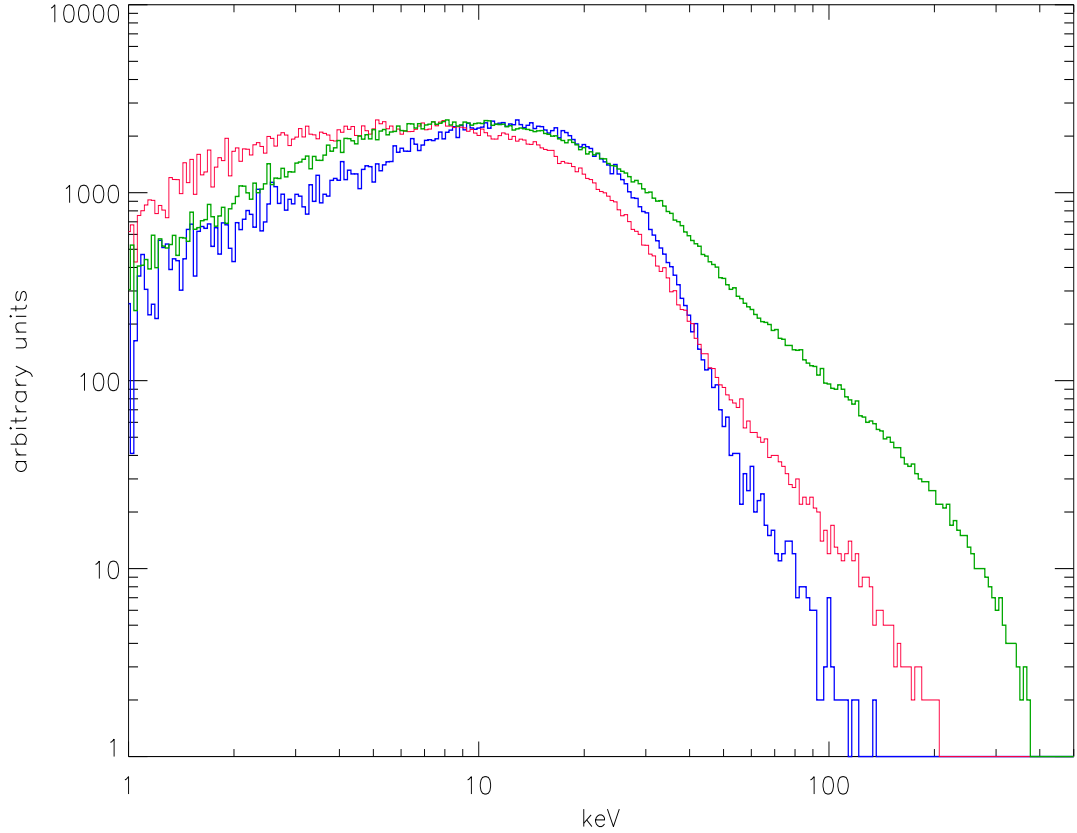}
      \caption{Emergent photon spectrum of the annihilated positrons. As the 511 keV emission from annihilation is produced at different radius from a BH,
      the line endure different redshift, thus producing a continuum. In these set of simulations,the temperature of the CC was 5 keV and its density
      was varied from $\dot m$ equal to 1 (blue line), 4 (red line), and 10 (green line). The main component of the spectra seems not to change much with $\dot m$, but a hard component rises when $\dot m$ increases. Due to the concurring effects of the pair creation efficiency and CC opacity to the 511 keV radiation, the spectrum is maximum for intermediate $\dot m=4$. The fit of the  $\dot m = 1$ spectrum with a blackbody spectral shape (kT = 6.2 keV) is shown by a dashed line.}
         \label{anni5}
   \end{figure}
%
   \begin{figure}
   \centering
   \includegraphics[width=15cm, height=12cm]{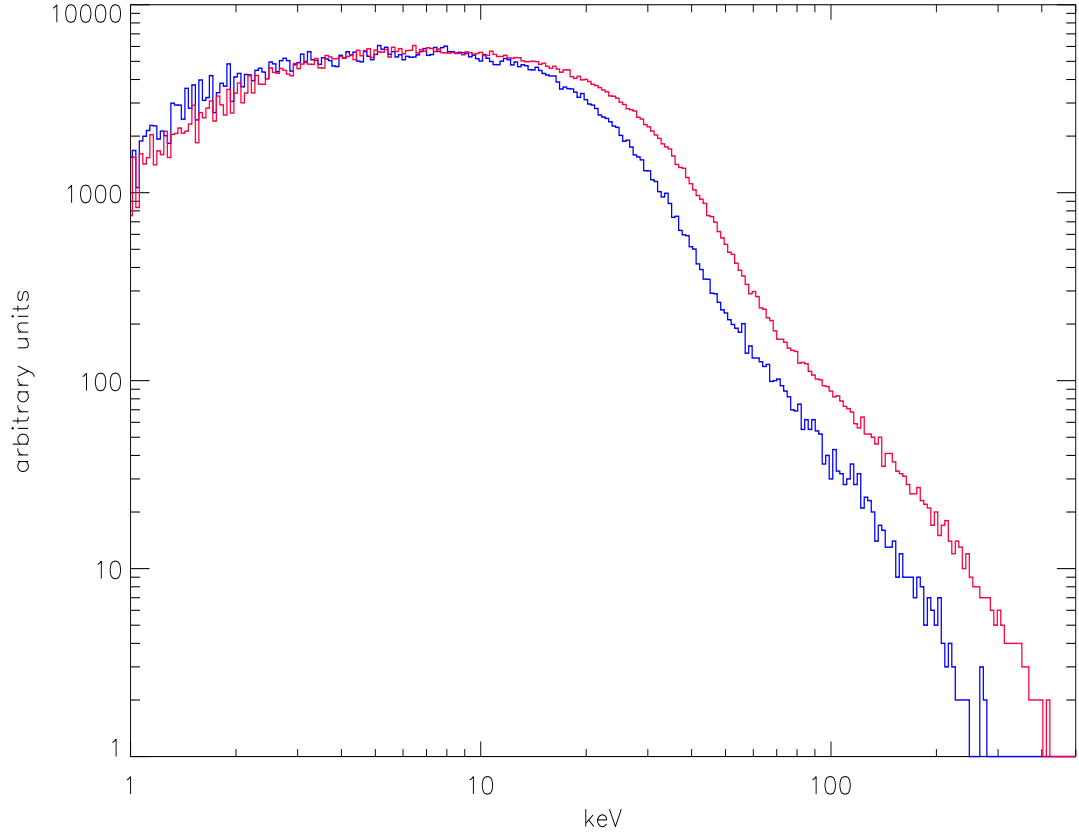}
      \caption{Emergent  photon spectrum of the annihilated positrons. In these set of simulations, 
     $\dot m$ equals to 4 and
    CC electron  temperatures are 5 keV (blue line) and 50 keV (red line). The cutoff energy of this annihilation line spectrum  only slightly changes with the cloud temperature. The fit of the  $\dot m = 4$ spectrum with a blackbody spectral shape (kT = 6.4 keV) is shown by a dashed line.}
         \label{anni50}
   \end{figure}
   
%

  \begin{figure}
   \centering
   \includegraphics[width=15cm, height=12cm]{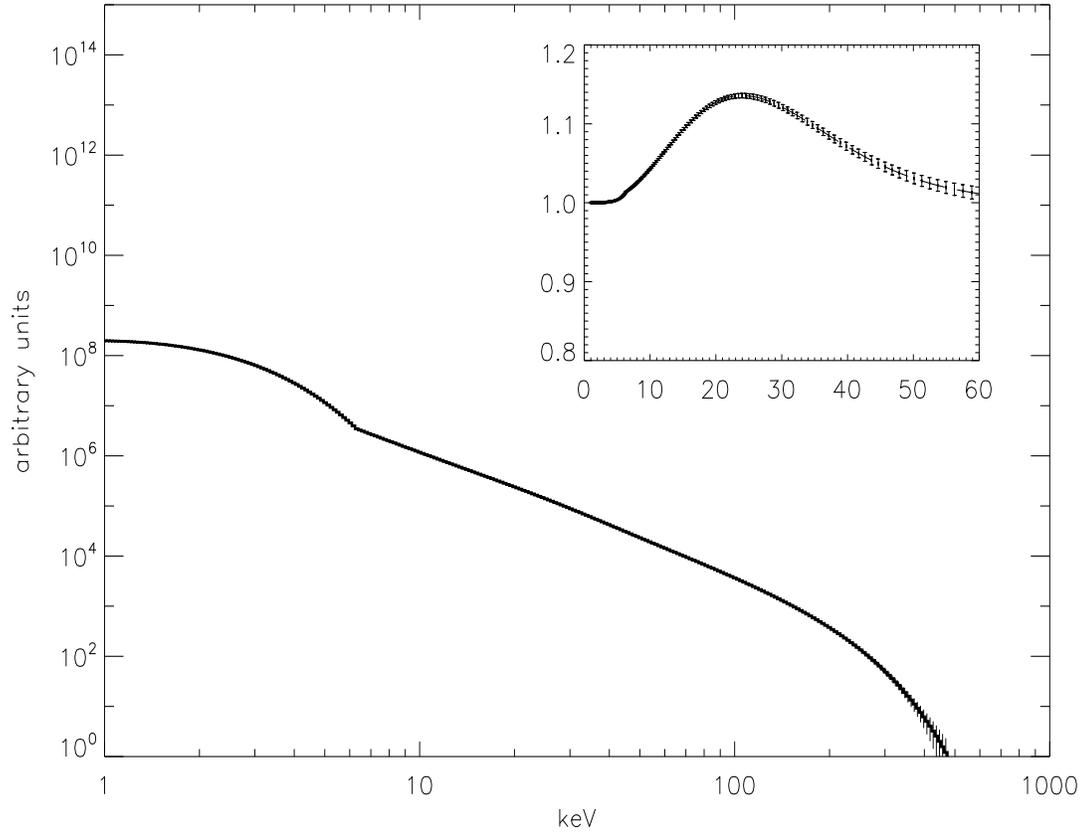}
      \caption{Emergent photon spectrum which includes a gravitationally redshifted  annihilated line.
      On the upper encapsulated right-hand panel we include the redshifted  annihilation line as a ratio of the resulting spectral values  to the Comptonized continuum.  }
         \label{anni5_total}
   \end{figure}

   \begin{figure}
   \centering
   \includegraphics[width=15cm, height=12cm]{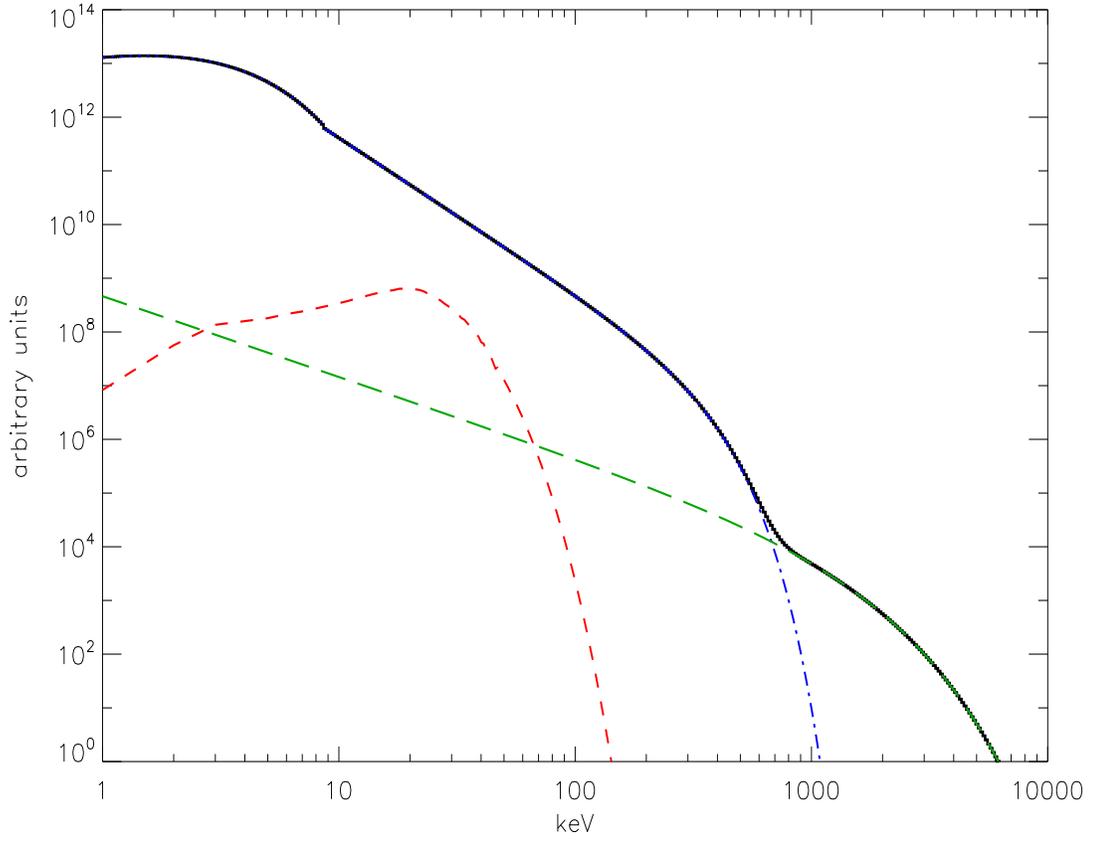}
      \caption{Spectrum at infinity resulting from the redshifted pair annihilation line (red) the Comptonization of X-ray photons on the CC electrons (blue) and Comptonization on created pairs (blue). In that case, the CC temperature is 5 keV ($\dot m = 4$).}
         \label{addspec}
   \end{figure}

%
   \begin{figure}
   \centering
   \includegraphics[width=15cm, height=12cm]{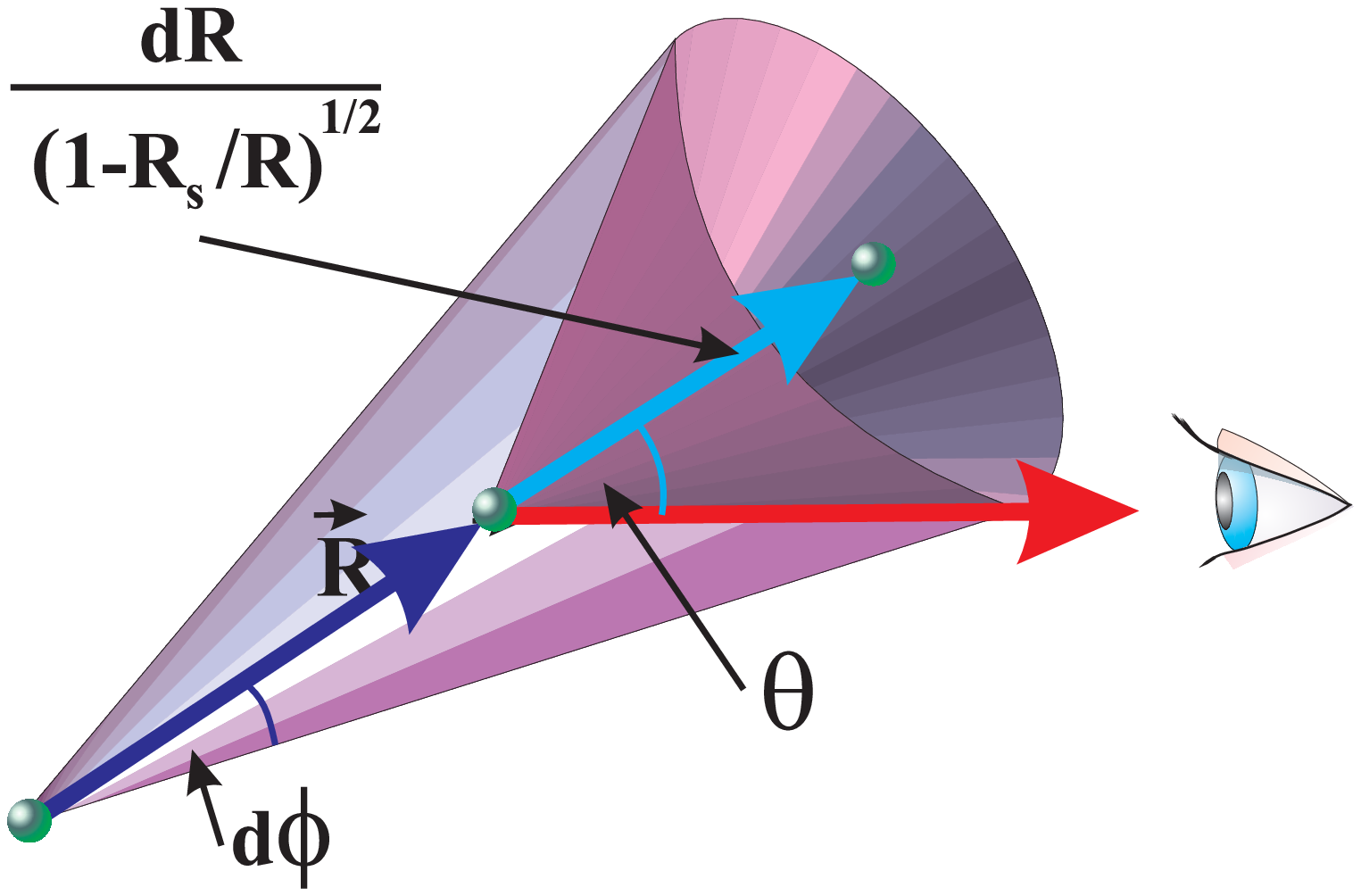}
      \caption{Sketch of light propagation in the Schwartzchild metric.}
         \label{lena}
   \end{figure}

%
   \begin{figure}
   \centering
   \includegraphics[width=10cm, height=10cm]{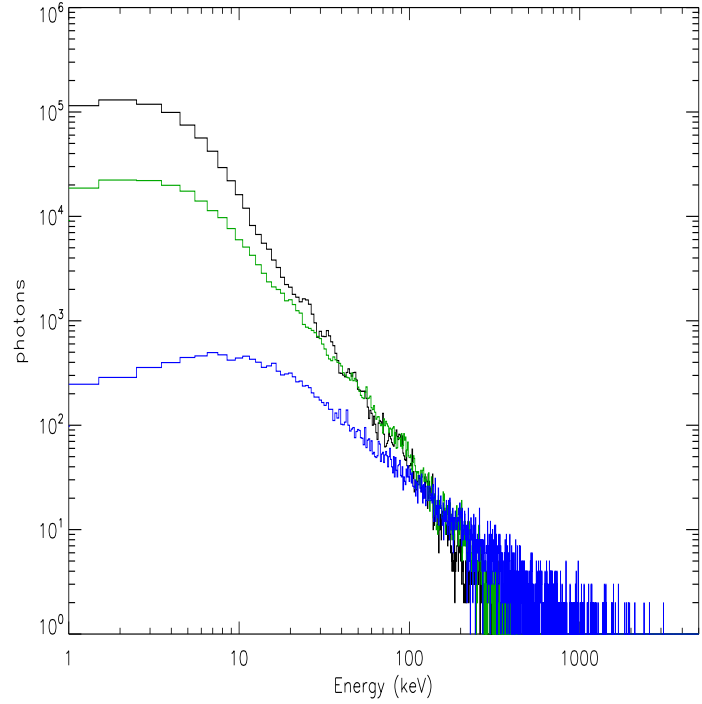}
      \caption{The photon spectra for three different radius ranges in the flow. 
The black  histogram is the emergent  spectrum seen by observers on Earth.
The green  histogram is the spectrum seen by observers staying at radial distance  $R=(1-1.1)R_{\rm S}$ from the black hole. 
The blue  histogram is the spectrum for observers staying at radial distance  $R=(1-1.01)R_{\rm S}$.
  }
         \label{spec_rad}
   \end{figure}
%
%
%
   \begin{figure}
   \centering
   \includegraphics[width=10cm, height=10cm]{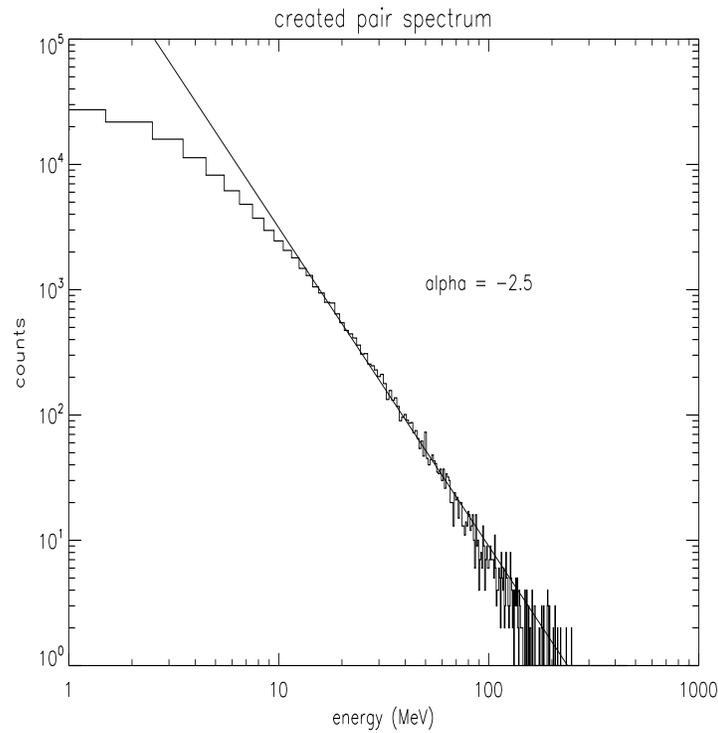} 
      \caption{ Spectrum of the created pairs. 
      Solid line is the best-fit of the simulated histogram by a power law which index is 2.5.
      }
         \label{EnergySpectrum}
   \end{figure}
%
   \begin{figure}
   \centering
   \includegraphics[width=10cm, height=10cm]{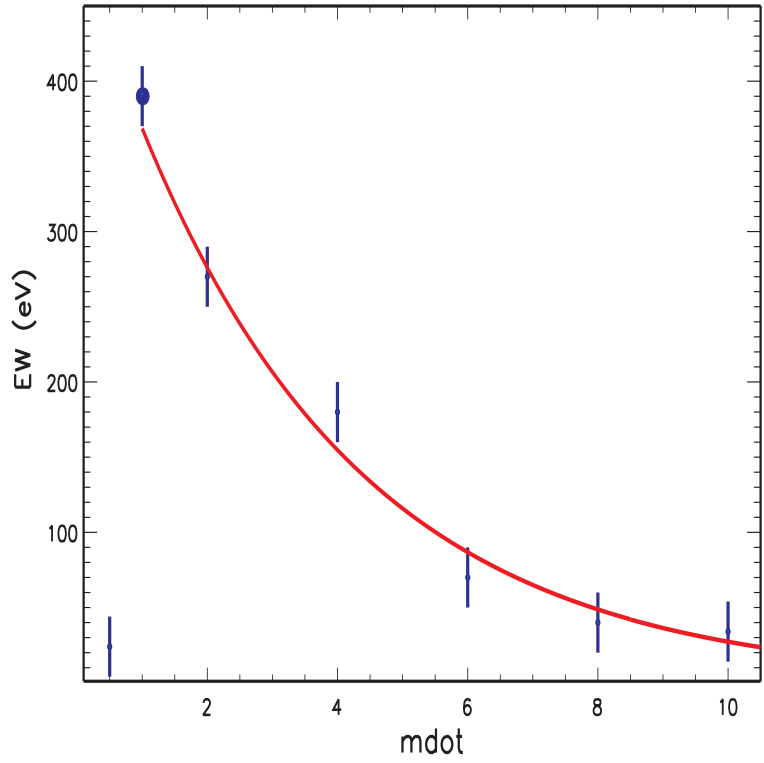} 
      \caption{ Evolution of the pair bump with the mass accretion rate $\dot m$. We measured the bump strength by computing its equivalent width (EW). Due to the concurring effects of the pair creation efficiency and CC opacity to the 511 keV radiation, the EW is maximum for a limited range of $\dot m$ (see text).
      }
         \label{EWmdot}
   \end{figure}
%

\end{document}